\documentstyle[11pt,iau212,twoside,epsf]{article}

\def\edcomment#1{\iffalse\marginpar{\raggedright\sl#1\/}\else\relax\fi}
\reversemarginpar

\begin{document}
\vspace*{1cm}
\title{Young open clusters in the Carina region}
 \author{Giovanni Carraro, Paola Marigo}
\affil{Dipartimento di Astronomia, Universit\`a di Padova, Vicolo Osservatorio
2, I-35122, Padova, Italy}
\author{Paolo Ventura}
\affil{Osservatorio Astronomico di Roma, Via di Frascati 33, I-00040, Monte
Porzio Catone, Italy}
\author{Martino Romaniello, Ferdinando Patat}
\affil{ESO, Karl Schwartzschild str. 2, D-85748, Garching bei Munchen, Germany}

\begin{abstract}
We present results of a new investigation (Carraro et al. 2002) aimed at 
clarifying the mutual relationship between the most prominent young open
clusters close to $\eta$ Carin\ae~, namely Trumpler~16, Trumpler~14 and Collinder~232.
\end{abstract}

\section{Introduction}
We have obtained deep UBVRI CCD photometry for Trumpler~14, Trumpler~16 and
Collinder~232, the three young open clusters closest to $\eta$ Carina\ae~. In 
this paper we briefly summarize our analysis methods and results. First of 
all we have tried to reconstruct the extinction pattern in the direction of 
the three clusters. By combining optical photometry with near infrared 
photometry and  spectral classification for a sample of stars in the three 
clusters, we find by using a variety of methods that the total to selective 
absorption ratio varies from cluster to cluster, being $R_V = 3.48\pm0.11, 4.16\pm0.07$, and $3.73\pm0.01$ for Trumpler~16, Trumpler~14 and Collinder~232, 
respectively. Then, by using the method devised by Romaniello et al. (2002),
we derived individual star color excesses and built up reddening corrected 
Colour-Magnitude Diagrams (CMDs), from which we constrain clusters's distances,
finding that Trumpler~14, Trumpler~16 and Collinder~232 are located 
$2.9\pm0.3$, $4.4\pm0.3$, and
$2.5\pm0.3$ kpc from the Sun, respectively. The method devised by Romaniello
et al. (2001) provides also stars luminosities and effective temperatures, that we used to construct  the HR diagram (HRD) for the clusters, which,
together with CMDs, have been used to derive  clusters' ages and age spreads.
To this aim pre-Main Sequence (MS) isochrones have been adopted from Ventura et al. (1998). We find that Trumpler~16 is older than Trumpler~14, and that all
the clusters have a prominent pre-MS population. As an example, in Fig.~1
we present our results for Collinder~232, that we conclude to be a real 
open cluster.

\begin{figure}
\plotone{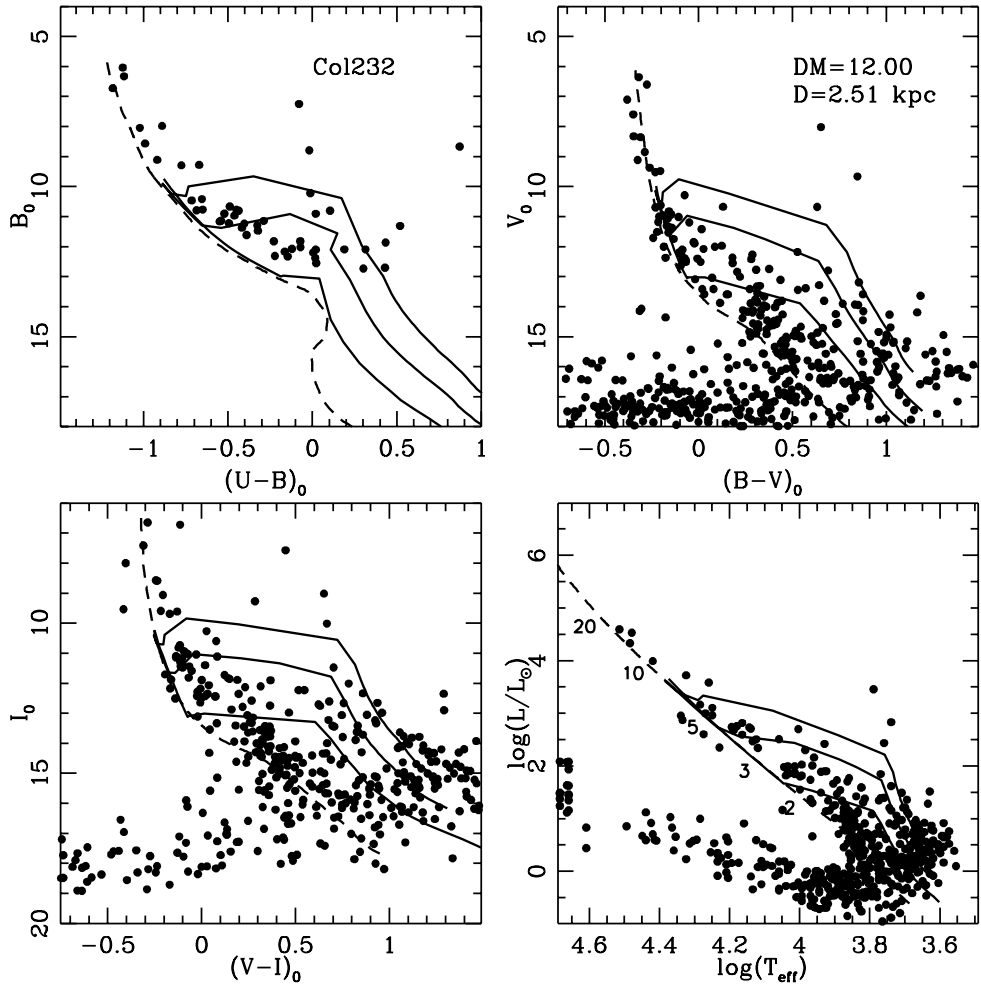}
\caption{CMDs and HRD for Collinder~232. The dashed line is and empirical ZAMS, the solid lines are pre-MS isocrones ($5\times 10^{5}$, $1$ and 
$5\times 10^{6}$)}
\end{figure}


\begin{references}
Carraro, G., et al. (2002), A\&A, submitted\\
Romaniello M., et al. (2002), AJ 123, 915\\
Ventura P., et al.1998, A\&A 334, 953\\
\end{references}
\end{document}